\documentclass[prl,preprint,amsmath]{revtex4-2}

\newcommand{\beq}{\begin{equation}}
\newcommand{\eeq}{\end{equation}}
\newcommand{\lab}{\label}

\newcommand{\taut}{\tilde \tau}
\newcommand{\sigmat}{\tilde \sigma}

\usepackage{graphicx}
\usepackage{xcolor}
\usepackage[normalem]{ulem}

\begin{document}

\title{Cluster Scaling and Critical Points: A Cautionary Tale}

\author{W.\ Klein}
\affiliation{Department of Physics, Boston University, Boston, Massachusetts 02215}
\affiliation{Center for Computational Science, Boston University, Boston, Massachusetts 02215}

\author{Harvey Gould}
\affiliation{Department of Physics, Boston University, Boston, Massachusetts 02215}
\affiliation{Department of Physics, Clark University, Worcester, Massachusetts 01610}

\author{Sakib Matin}
\affiliation{Department of Physics, Boston University, Boston, Massachusetts 02215}
\affiliation{Theoretical Division and CNLS, Los Alamos National Laboratory, Los Alamos, New Mexico 87546} 

\begin{abstract}

Many systems in nature are conjectured to exist at a critical point, including the brain and earthquake faults. The primary reason for this conjecture is that the distribution of clusters (avalanches of firing neurons in the brain or regions of slip in earthquake faults) can be described by a power law. Because there are other mechanisms such as $1/f$ noise that can produce power laws, other criteria that the cluster critical exponents must satisfy can be used to conclude whether or not the observed power law behavior indicates an underlying critical point rather than an alternate mechanism. We show how a possible misinterpretation of the cluster scaling data can lead to incorrectly conclude that the measured critical exponents do not satisfy these criteria. Examples of the possible misinterpretation of the data for one-dimensional random site percolation and the one-dimensional Ising model are presented. We stress that the interpretation of a power law cluster distribution indicating the presence of a critical point is subtle, and its misinterpretation might lead to the abandonment of a promising area of research.
\end{abstract}

\maketitle

Cluster scaling techniques, such as the determination of the exponents which characterize the distribution of clusters and the scaling laws which relate the different exponents, have been used as an indicator that a system is operating at a critical point. This approach have been applied to many systems, including earthquake faults~\cite{rundle-rev, agubook}, the brain~\cite{beggs,kinouchi2006optimal, meisel2013fading, yu2013universal, Kessenich2016Synaptic, hahn2017spontaneous, munoz2018RevMod, Fontele_PRL19, levina2019Critical, alain} and micelles~\cite{mic, biol}. In this work, we discuss the subtleties of applying cluster scaling as a test for critical points, especially in neural systems. We highlight common pitfalls when applying cluster scaling to two exactly solved models, namely the one-dimensional (1D) site-percolation and the 1D Ising models, and discuss how to avoid them. 

The idea of cluster scaling was formulated in its most quantitative form in the context of percolation~\cite{AS}. A percolation model is specified by defining the objects to be connected, the rules which define the connections, and how the objects are distributed (such as random or correlated). Percolation models often exhibit a phase transition as a function of a parameter such as the probability of an occupied site or bond~\cite{AS}. The percolation transition usually involves the appearance of an infinite or spanning cluster~\cite{AS}. 

In the neighborhood of the percolation transition, the probability $P(s)$ of a cluster of size $s$ can be described by the critical exponents $\tau$ and $\sigma$ in the Fisher-Stauffer scaling relation~\cite{AS}
\beq
\lab{FS}
P(s) \sim s^{-\tau + 1} e^{-s/s_c} \mbox{ and } s_c \sim \epsilon^{-1/\sigma} \quad (s \gg 1),
\eeq
where $s_c(\epsilon)$ is the characteristic size of the clusters. The parameter $\epsilon$ is the relevant scaling field and is the independent tuning parameter in the system. Here, $\epsilon = 0$ corresponds to the percolation critical point.

The number of clusters with duration $D$ satisfies a scaling relation similar to Eq.~\eqref{FS} with $\tau$ replaced by $\tau_D$ (and $\sigma$ replaced by $\sigma_D$)~\cite{neural-neutral-prx, sakib-neuron, sethna}. The critical behavior at the percolation transition is analogous to the behavior at thermal critical points~\cite{AS}, which are characterized by long-range correlations, a divergent response to external stimuli, and fluctuations at all scales~\cite{AS}. The existence of cluster scaling can also indicate the presence of an underlying thermal critical point as is found in the Ising model~\cite{coniglio-etal,ck}.

The question of whether the existence of cluster scaling actually indicates the existence of an underlying percolation
and/or 
thermal critical point can be subtle~\cite{agubook, beggs} because several mechanisms can result in scaling that is not associated with a critical point~\cite{beggs, alain}. A necessary but perhaps insufficient test of whether cluster scaling implies that there is an underlying critical point is that the critical exponents $\nu$ and $z$ are related to the cluster exponents $\tau$, $\sigma$, and $\tau_D$ by the relation~\cite{dahmen} 
\beq
\lab{necessary}
\frac{\tau_D- 1}{\tau - 1} = \frac{1}{\sigma\nu z}.
\eeq
The exponent $\nu$ characterizes the divergence as $\epsilon\to 0$ of the connectedness and/or the correlation length $\xi\sim \epsilon^{-\nu}$~\cite{AS}, and the dynamical exponent $z$ characterizes the divergence of the the correlation time $\tau\sim \xi^{z}$~\cite{dahmen}.

When cluster scaling is applied to models of the brain or earthquake faults, the cluster may correspond to the number of neurons that have fired or the area that has slipped in an earthquake~\cite{serino}. Touboul and Destexhe~\cite{alain} have recently argued that the experimentally determined exponents associated with the scaling of neural avalanches 
do not satisfy Eq.~\eqref{necessary}. Hence, they conclude that the scaling of clusters of fired neurons does not indicate that the brain is operating at a critical point. 
In the following we argue that their conclusion may be based on a possible misinterpretation of the cluster scaling data, and that further investigation is needed before it can be determined whether or not the observed cluster scaling is associated with an underlying critical point.

We first discuss cluster scaling for 1D site percolation. The probability of a cluster of size $s$ is given by Eq.~\eqref{FS}, with $p$, the probability that a site is occupied, $q = 1 - p$, and $s_c \sim q^{-1/\sigma}$. The percolation exponent that characterizes the divergence of the mean cluster size $\gamma_{p}$ is related to $\tau$ and $\sigma$ by
\begin{align}
\lab{gamma}
\gamma_{p} & = \frac{3 - \tau}{\sigma}, \\
\noalign{\noindent and the order parameter exponent $\beta_{p}$ is given by}
\lab{beta} 
\beta_{p} & = \frac{\tau - 2}{\sigma}.
\end{align}
The relations~\eqref{gamma} and \eqref{beta} can be obtained by evaluating the first and second moments of $P(s)$ in Eq.~\eqref{FS} and taking the argument of the exponential to be zero up to $qs^{\sigma} \sim 1$, resulting in integrals of the power law part of $P(s)$ from $s=1$ to $s=1/q^{1/\sigma}$. In $d=1$, the exponents $\gamma_{p} = 1$ and $\beta_{p} = 0$ are known exactly~\cite{RSK}, which implies that $\tau = 2$ and $\sigma = 1$. 

\begin{figure}[t]
\includegraphics[scale=0.48]{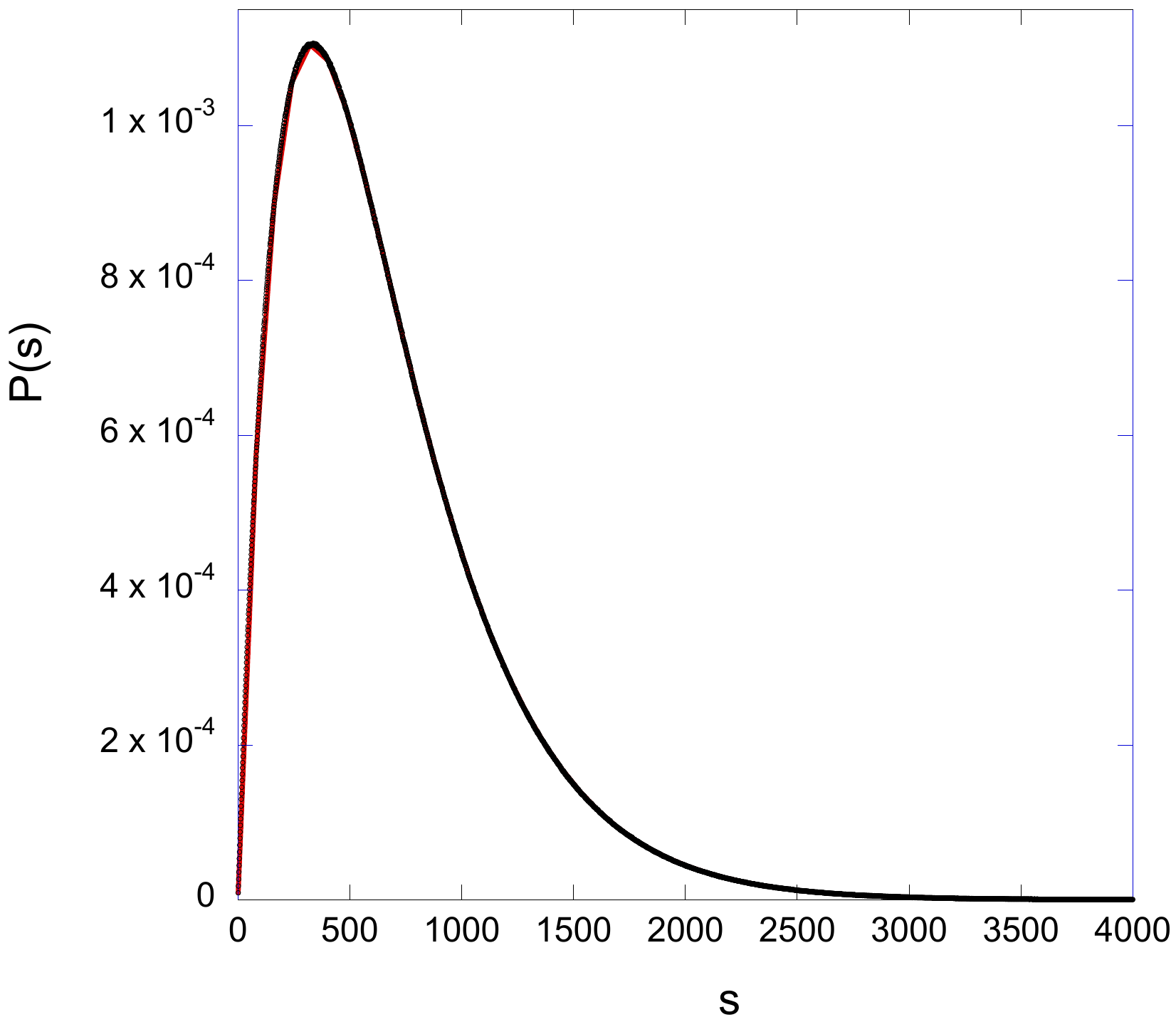}
\includegraphics[scale=0.48]{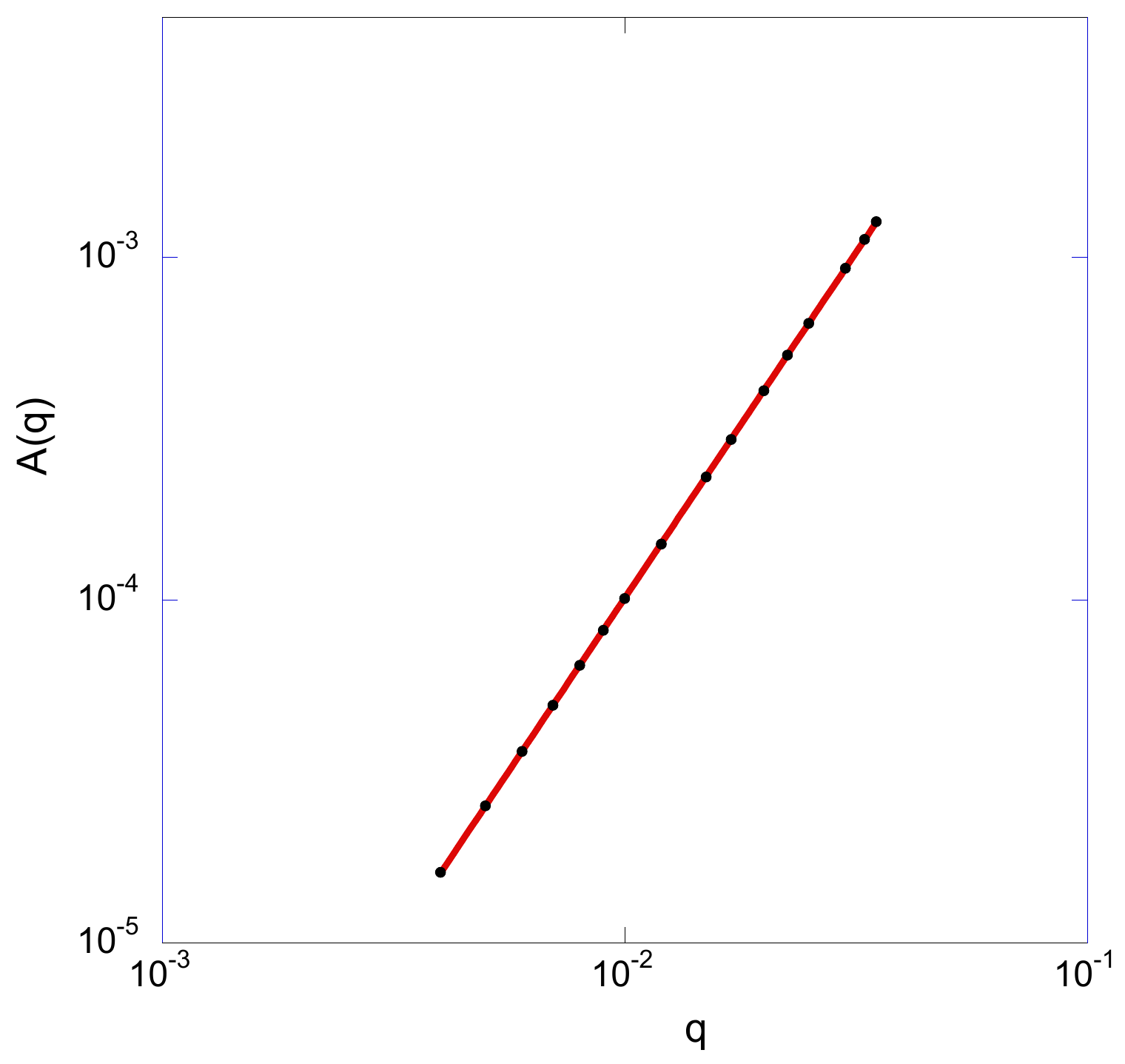}
\vspace{-0.2in}
\caption{\label{fig:Psite}(a) The probability $P(s)$ of a cluster of $s$ sites grown from a seed using the Leath algorithm~\cite{leath} for a 1D lattice of length $L=20001$ with $p=0.997$ and $10^9$ trials. $P(s)$ is well approximated by $A s\, e^{-b s}$, with $A \approx 9.03 \times 10^{-6}$ and $b \approx 3.0 \times 10^{-3}$, corresponding to $\taut_p - 1 = 0$ and $\sigmat_p = 1$. (b) The quadratic dependence on $q = 1 - p$ of the amplitude $A(q)$ of the simulation results.}
\end{figure}

Figure~\ref{fig:Psite}(a) shows the results of a simulation of the probability $P(s)$ of a cluster of size $s$ grown from a seed using the Leath algorithm~\cite{leath}. To avoid confusion, we denote the exponents obtained from the simulations by a tilde. We see that $P(s)$ is well approximated by the function $s e^{-b s}$, corresponding to $\taut_p - 1 =0$ and $\sigmat_p = 1$. If we use these values of $\taut$ and $\sigmat$ in Eqs.~\eqref{gamma} and \eqref{beta}, we would find $\gamma = 2$ and $\beta = -2$, which disagrees with the exact results.

To understand why the simulation does not yield the theoretical value of $\tau=2$, we calculate the probability of a cluster of $s$ sites exactly. We choose a seed site at random and occupy it with probability $p$. We then occupy its two neighbors with probability $p$. If we generate two empty sites with probability $q^{2}$, the cluster growth is terminated. Hence, the probability of a cluster with $s$ sites is given by 
\beq
\lab{1-d prob}
P(s) = s q^{2} p^{s} + q\delta_{s,0},
\eeq
where the factor of $s$ multiplying $q^{2}p^{s}$ is due to the fact that any of the $s$ sites in the cluster could be the seed~\cite{AS}; $\delta_{s,0}$ is the Kronecker delta function. It is easy to show that $P(s)$ is normalized. There is a percolation phase transition at $p=1$~\cite{RSK}.

We can express $P(s)$ in the Fisher-Stauffer form of Eq.~\eqref{FS} by writing 
\beq
\lab{FS-1d-perc}
P(s) = s\, q^{2} e^{s\ln p} \approx q^{2}s\, e^{-qs},
\eeq
where $q \ll 1$, and we have ignored the $\delta_{s,0}$ term because the exponents are determined by large clusters. The probability distribution in Eq.~\eqref{FS-1d-perc} can be used to calculate $\beta_{p}$ and $\gamma_{p}$ with the expected results $\beta_p = 0$ and $\gamma = 1$. However, if we compare $P(s)$ in Eq.~\eqref{FS-1d-perc} for a {\it fixed value} of $q$ to Eq.~\eqref{FS}, we would conclude that $\tau - 1 = 0$ (and $\sigma = 1$), consistent with the simulation result in Fig.~\ref{fig:Psite}(a), but inconsistent with the exact result $\tau=2$.

Given the exact solution for $P(s)$ in Eq.~\eqref{FS-1d-perc}, the resolution of this apparent contradiction is clear. Because the value of $q$ was fixed in the simulations, the factor of $q^{2}$ in Eq.~\eqref{FS-1d-perc} was not accounted for when obtaining the value of $\tau$ to use in Eqs.~\eqref{gamma} and \eqref{beta}. Hence, for 1D site percolation, the correct value of the critical exponent $\tau$ cannot be obtained from a measurement of the cluster distribution at a fixed value of $q$.

To determine the correct value of $\tau$, we can reverse the scaling argument used to determine Eqs.~\eqref{gamma} and \eqref{beta} and use the fact that the resultant integrals will be evaluated at $q s^{\sigma}=1$. Hence, we can replace $q$ by $1/s$ in Eq.~\eqref{FS-1d-perc} and write 
\beq
\lab{FS-1d-perc2}
P(s) \sim \frac{s e^{-q s}} {s^{2}} \qquad (s \gg 1).
\eeq
From the form of $P(s)$ in Eq.~\eqref{FS-1d-perc2}, we immediately determine the correct value $\tau = 2$. The quadratic dependence of the amplitude of the cluster distribution on the scaling parameter $q$ determined from the simulations in shown in Fig.~\ref{fig:Psite}(b).

We stress that the form of $P(s)$ in Eq.~\eqref{FS-1d-perc} is the same as would be found if the cluster distribution were measured {\it for one value of $q$} (the relevant scaling field). We cannot measure the correct value of $\tau$ from $P(s)$ from a single value of $q$ if we do not correctly account for the $q^2$ dependence of the amplitude. 

We have seen that the measurement of the $\tau$ exponent from the probability of finding a cluster of size $s$ for one value of the scaling field can lead to missing the relation between cluster scaling and a possible underlying percolation-like  
critical point.

To illustrate the relation between cluster scaling and a thermal critical point, we consider the temperature $T = 0$ critical point of the 1D nearest-neighbor ferromagnetic Ising model in zero magnetic field. The Ising correlation length exponent is $\nu=1$~\cite{stanley} and the dynamical critical exponent is $z=2$ for model A dynamics~\cite{hh, tobochnik, stanley}.

We can map the Ising critical point onto a percolation transition by assigning a down spin to be an occupied site and an up spin to be empty. The distribution of spins satisfies the Boltzmann probability distribution associated with the Ising Hamiltonian. We assign a percolation bond between nearest-neighbor occupied sites with probability $p_{b} = 1 - e^{-2K} = 1- q_{b}$, where $K = J/k_{B}T$, $J$ is the coupling constant, and $k_{B}$ is the Boltzmann constant~\cite{ck}. The thermal problem can be treated as a distribution of independent Ising bonds, with the probability of an Ising bond given by~\cite{huang}
\beq
\lab{bond-prob-I}
p_{I} = \frac{e^K}{e^K + e^{-K}} \approx 1 - e^{-2K} = 1 - q_{I} \qquad (K \gg 1).
\eeq

We generate a cluster by choosing a down spin seed site and adding a site to the cluster if both the Ising and percolation bonds are present, which occurs with probability $p=p_{I}p_{b}$. The cluster grows until an empty bond is encountered on both sides of the cluster with probability $(q_{b} + q_{I})^2$. Hence, the probability $P(s)$ of an Ising cluster of size $s$ is
\beq
\lab{cl-I}
P(s) = s(q_{b} + q_{I})^{2}(p_{I}p_{b})^{s}.
\eeq 
For $K \gg 1$ we have $p_{b} = p_{I}$ and $q_{b} = q_{I}$, and
\beq
\lab{p=p} 
p \equiv p_b p_I = (1 - e^{-2K})^{2} \sim 1 - 2e^{-2K} = 1 - (q_{b} + q_{I}) = 1 - 2e^{-2K} = 1 - q.
\eeq
Therefore, the probability of a cluster of size $s$ near the $T=0$ critical point is 
\beq
\lab{cl-I-m}
P(s) = s q^{2} p^{s},
\eeq
which is identical to Eq.~\eqref{1-d prob}. Because the form of Eq.~\eqref{cl-I-m} is the same as for 1D random site percolation, we conclude that $\tau$ and $\sigma$ in Eq.~\eqref{necessary} for the 1D Ising model are the same as the 1D random percolation exponents~\cite{RSK}.

\begin{figure}[h]
 \includegraphics[scale=0.54]{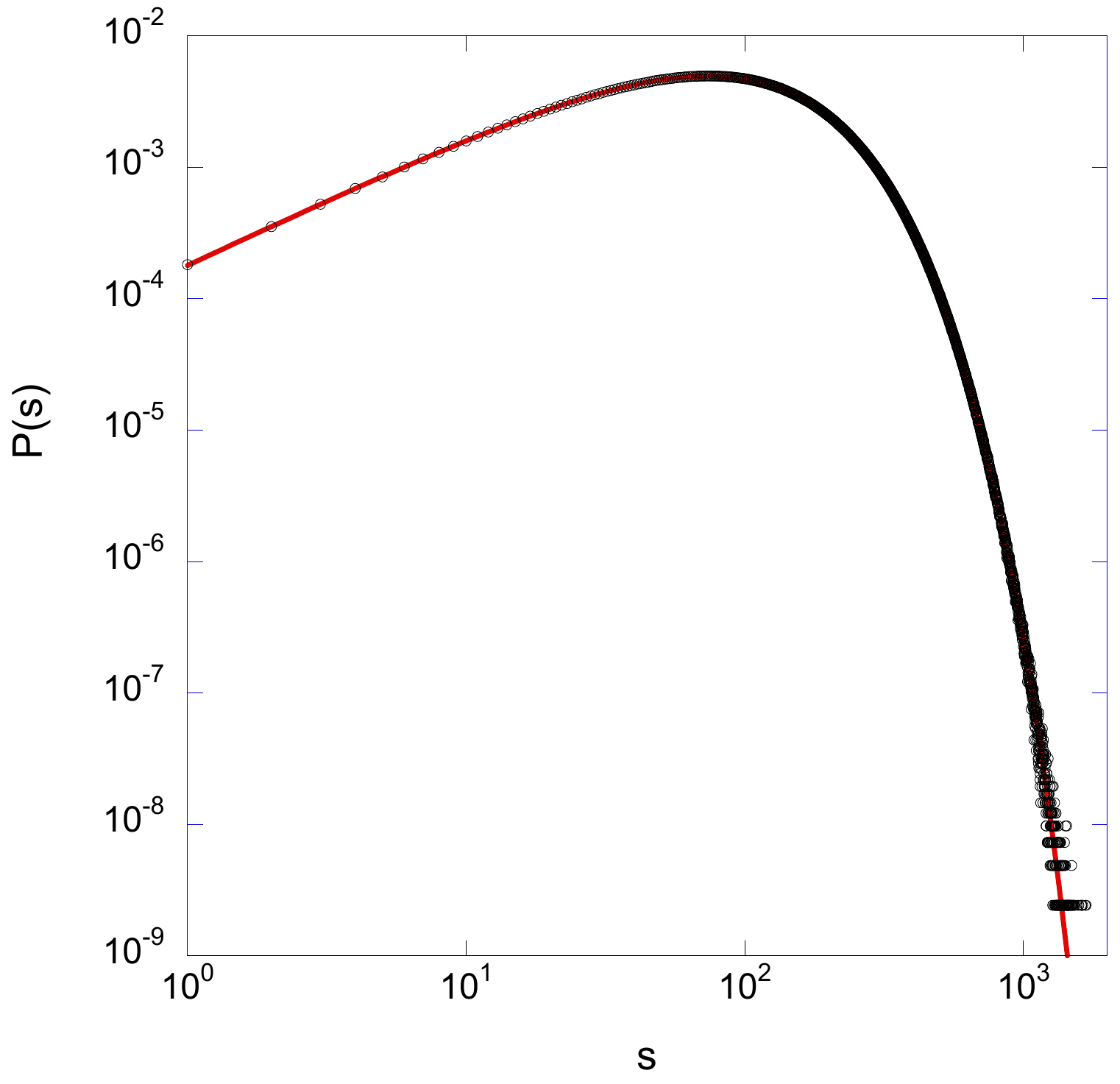}
\vspace{-0.2in}
\caption{\label{fig:ising}(a) The probability $P(s)$ of a cluster of $s$ spins for the 1D Ising model at temperature $T=0.4$ with periodic boundary conditions and $L=10^5$ averaged over $10^6$ Monte Carlo steps per spin. The form of $P_s$ is consistent with $A s e^{-b s}$, with $A \approx 1.82 \times 10^{-4}$ and $b \approx 1.35 \times 10^{-2}$, corresponding to $\taut - 1 = 0$ and $\tilde \sigma = 1$ in Eq.~\eqref{FS}.}
\end{figure}

The dynamics of the Ising model is characterized by the random walk dynamics of the domain walls. To determine the duration exponent $\tau_D$, we note that the ``distance'' a domain wall walker must travel in one dimension is $s$ and $s$ is the correlation length. The probability that there is a cluster of size $s$ at the critical point is 
\beq
\lab{prob-S-crit}
P(s) \sim \frac{1}{s},
\eeq
where we have used $q^{2} \sim 1/s^{2}$. Because the lifetime $D$ of the cluster is proportional to $s^{2}$, the probability that there is a cluster with lifetime $D$ at the critical point is
\beq
\lab{prob-t-crit}
P(D) \sim \frac{1}{(s^{2})^{1/2}} = \frac{1}{D^{1/2}},
\eeq
which implies that $\tau_{D} - 1 = 1/2$ or $\tau_D = 3/2$. 

We substitute the values $\tau_{D} - 1 = 1/2$, $\tau - 1 = 1$, $\sigma = 1$, $\nu = 1$ and $z = 2$ in Eq.~\eqref{necessary} and find that this consistency condition is satisfied. This result is not surprising because we constructed the percolation model to be isomorphic to the Ising critical point~\cite{ck}. In contrast, the simulation results in Fig.~\ref{fig:ising} for the probability $P(s)$ of a cluster of $s$ spins for the $d = 1$ Ising cluster probability at one value of the temperature yield an exponential for $s \gg 1$, or $\taut - 1 = 0$, which would imply that $\taut_{D} - 1 = 0$ from the random walk argument, rather than the theoretical result $\tau_D = 3/2$. Hence, the ratio of $\tau_{D}-1$ to $\tau - 1$ would be 1, and because $z$, $\nu$, and $\sigma$ remain the same, the consistency condition in Eq.~\eqref{necessary} would not be satisfied, leading to an erroneous conclusion. Note that although $P(s)$ for the Ising model is not a power law distribution, the behavior of $P(s)$ corresponds to a critical point.

We note that the prefactor of $P(s)$ would not vanish as the critical point is approached if the system self-organizes to the critical point. Hence, the failure of the cluster critical exponents to satisfy the consistency condition in Eq.~\eqref{necessary} for a single measurement appears to rule out the possibility that the underlying critical point (if it exists) is self-organized. 
Nonetheless, there remains the possibility that a different definition of the clusters might satisfy Eq.~\eqref{necessary} with a nonvanishing prefactor. Such a situation would be analogous to the case of the $d=2$ nearest-neighbor Ising model, which requires a more subtle definition of the clusters associated with the underlying thermal critical point~\cite{ck}.

Finally, we note that the examples of the vanishing of the cluster scaling prefactor we have discussed are for models in which both the cluster and thermal order parameter exponent $\beta$ is zero. In future work, we intend to investigate whether this connection is general, and if so, its physical basis. 

In summary, our results illustrate that Eq.~\eqref{necessary} must be used with caution and that the exponents associated with cluster scaling at one value of the scaling parameter may not correctly determine the existence of an underlying critical point. Instead, a true test requires that the cluster distribution be measured at several values of the scaling field. Unfortunately, such measurements may not be possible for physical systems such as earthquake faults and in vivo neural systems. We also stress it is important that a possible misinterpretation of the data should not prematurely lead to the abandonment of a promising direction of research.

\begin{acknowledgments}
We thank Karin A.\ Dahmen and Jan Tobochnik for useful conversations.
\end{acknowledgments}


\begin{thebibliography}{99}

\bibitem{rundle-rev} J. B. Rundle, S. Stein, A. Donnellan, D. L. Turcotte, W. Klein, and C. Saylor, ``The complex dynamics of earthquake fault systems: new approaches to forecasting and nowcasting of earthquakes,'' Rep. Prog. Phys. {\bf 81}, 076801 (2021).

\bibitem{agubook} {\sl Geo-complexity and the Physics of Earthquakes}, edited by J. B. Rundle, D. L. Turcotte and W. Klein (American Geophysical Union, Washington, DC, 2000).

\bibitem{beggs} J. M. Beggs, {\sl The Cortex and the Critical Point} (MIT Press, 2022).

\bibitem{kinouchi2006optimal}O. Kinouchi and M. Copelli, 
``Optimal dynamical range of excitable networks at criticality,''
Nature Physics {\bf 2}, 348 (2006).
 
\bibitem{meisel2013fading}
C. Meisel, E. Olbrich, O. Shriki, and P. Achermann, 
``Fading signatures of critical brain dynamics during sustained wakefulness in humans,'' J. Neuroscience {\bf 33}, 17363 (2013).

\bibitem{yu2013universal}S. Yu, H. Yang, O. Shriki, and D. Plenz, 
``Universal organization of resting brain activity at the thermodynamic critical point,'' Frontiers in Systems Neuroscience {\bf 7}, 42 (2013).

\bibitem{Kessenich2016Synaptic}L. Michiels Van Kessenich, L. de Arcangelis, and H. J. Herrmann, ``Synaptic plasticity and neuronal refractory time cause scaling behaviour of neuronal avalanches,'' Scientific Reports {\bf 6}, 32071 (2016).

\bibitem{hahn2017spontaneous}G. Hahn et al., 
``Spontaneous cortical activity is transiently poised close to criticality,''
PLoS Computational Biology {\bf 13}, e1005543 (2017).

\bibitem{munoz2018RevMod}M. A. Mu\~noz, ``Colloquium: Criticality and dynamical scaling in living systems,'' Rev. Mod. Phys. {\bf 90}, 031001 (2018).

\bibitem{Fontele_PRL19}A. J. Fontenele et al.,
``Criticality between cortical states,'' Phys. Rev. Lett. {\bf 122}, 208101 (2019).

\bibitem{levina2019Critical}A. Das and A. Levina,
``Critical neuronal models with relaxed timescale separation,''
Phys. Rev. X {\bf 9}, 021062 (2019).
 

\bibitem{alain} J. Touboul and A. Destexhe, ``Power-law statistics and universal scaling in the absence of criticality,'' Phys. Rev. E {\bf 95}, 012413 (2017) and references therein.

\bibitem{mic} F. Mallamace et al.,
``Percolation and critical phenomena of an attractive micellular system,'' Fractals {\bf 11}, 37 (2003).

\bibitem{biol} P. Nelson, {\sl Physical Models of Living Systems}, 2nd ed. (Chiliagon Science, 2021).

\bibitem{AS} A. Aharony and D. Stauffer, {\sl Introduction to Percolation Theory}, 2nd ed. (Taylor \& Francis, 1994).

\bibitem{sethna} J. P. Sethna, K. A. Dahmen, and C. R. Meyers, ``Crackling noise,'' Nature {\bf 410}, 242 (2001).

\bibitem{sakib-neuron} S. Matin, T. Tenzin, and W. Klein, ``Scaling of causal neural avalanches in a neutral model,'' Phys. Rev. Res. {\bf 3}, 013107 (2021).

\bibitem{neural-neutral-prx} Matteo Martinello, Jorge Hidalgo, Amos Maritan, Serena di Santo, Dietmar Plenz, and Miguel A. Munoz, ``Neutral theory and scale-free neural dynamics,'' Phys. Rev. X. {\bf 7}, 041071 (2017).

\bibitem{coniglio-etal} A. Coniglio, C. Nappi, F. Perrugi and L. Russo, ``Percolation and phase transitions in the Ising model,'' J. Phys. A Math. Gen. {\bf 13}, 1783 (1977).

\bibitem{ck} A. Coniglio and W. Klein, ``Correlated site-bond percolation and Ising critical droplets: a renormalization group approach,'' J. Phys. A {\bf 13}, 2775 (1980).

\bibitem{dahmen} J. P. Coleman, K. A. Dahmen, and R. L. Weaver, ``Avalanches and scaling collapse in the large-$N$ Kuramato model,'' Phys. Rev. E {\bf 97}, 042219 (2018).

\bibitem{serino} C. Serino, K. F. Tiampo and W. Klein, ``A new approach to Gutenburg-Richter scaling,'' Phys. Rev. Lett. {\bf 106}, 108501 (2011).

\bibitem{RSK} P. J. Reynolds, H. E. Stanley, and W. Klein, ``Ghost
fields, pair connectedness, and scaling: Exact results in one
dimension,'' J. Phys. A {\bf 10}, L203 (1977).

\bibitem{leath} P. L. Leath, ``Single cluster growth,'' Phys. Rev. B {\bf 14}, 5046 (1976).

\bibitem{stanley} H. E. Stanley, {\sl Introduction to Phase Transitions and Critical Phenomena} (Oxford University Press, Oxford, 1971).

\bibitem{hh}P. C. Hohenberg and B. I. Halperin, ``Theory of dynamic critical phenomena,'' Rev. Mod. Phys. {\bf 49}, 435 (1977).

\bibitem{tobochnik}R. Cordery, S. Sarker, and J. Tobochnik, ``Physics of the dynamical critical exponent in one dimension,'' Phys. Rev. B {\bf 24}, 5402(R) (1981).

\bibitem{huang}K. Huang, {\sl Statistical Mechanics}, 2nd ed. (John Wiley \& Sons, 1987).

\bibitem{OFC} Z. Olami, H. J. S. Feder, and and K. Christensen, ``Self-organized criticality in a continuous, nonconservative cellular automaton modeling earthquakes," Phys. Rev. Lett. {\bf 68}, 1244 (1992).

\bibitem{RJ} J. B. Rundle and D. D. Jackson, ``Numerical simulation of earthquake sequences,'' Bull. Seism. Soc. Am. {\bf 67}. 1363 (1977).

\bibitem{kgml} W. Klein, H. Gould, S. Matin, and N. Leung, ``Equilibrium theory of a driven dissipative system'' (in preparation).

\end{thebibliography}
\end{document}